\documentclass[prb,twocolumn,showpacs,preprintnumbers,amsmath,amssymb,superscriptaddress]{revtex4}

\usepackage{graphicx}
\usepackage{dcolumn}
\usepackage{bm}

\begin{document}

\preprint{APS/123-QED}

\title{Radiative quantum efficiency in an InAs/AlSb intersubband transition}

\author{C. Faugeras}
\affiliation{Mat\'eriaux et Ph\'enom\`enes Quantiques,
Universit\'e Paris 7, 75251 Paris Cedex 05, France}
\author{A. Wade}
\affiliation{National High Magnetic Field Laboratory
 1800 E. Paul Dirac Dr. Tallahassee, FL 32310-3706}
\author{A. Leuliet}
\affiliation{Mat\'eriaux et Ph\'enom\`enes Quantiques,
Universit\'e Paris 7, 75251 Paris Cedex 05, France}
\affiliation{Thales Research and Technology, Route departementale
128, 91767 Palaiseau cedex, France}
\author{A. Vasanelli}
\affiliation{Mat\'eriaux et Ph\'enom\`enes Quantiques,
Universit\'e Paris 7, 75251 Paris Cedex 05, France}
\author{C. Sirtori}
\email{carlo.sirtori@paris7.jussieu.fr} \affiliation{Mat\'eriaux
et Ph\'enom\`enes Quantiques, Universit\'e Paris 7, 75251 Paris
Cedex 05, France}
\author{G. Fedorov}
\affiliation{National High Magnetic Field Laboratory
 1800 E. Paul Dirac Dr. Tallahassee, FL 32310-3706}
\author{D. Smirnov}
\affiliation{National High Magnetic Field Laboratory
 1800 E. Paul Dirac Dr. Tallahassee, FL 32310-3706}
\author{R. Teissier}
\affiliation{Centre d'Electronique et de Micro-opto\'electronique
de Montpellier, UMR 5507 CNRS - Universit\'e Montpellier II, 34095
Montpellier Cedex 05, France}
\author{A.N. Baranov}
\affiliation{Centre d'Electronique et de Micro-opto\'electronique
de Montpellier, UMR 5507 CNRS - Universit\'e Montpellier II, 34095
Montpellier Cedex 05, France}
\author{D. Barate}
\affiliation{Centre d'Electronique et de Micro-opto\'electronique
de Montpellier, UMR 5507 CNRS - Universit\'e Montpellier II, 34095
Montpellier Cedex 05, France}
\author{J. Devenson}
\affiliation{Centre d'Electronique et de Micro-opto\'electronique
de Montpellier, UMR 5507 CNRS - Universit\'e Montpellier II, 34095
Montpellier Cedex 05, France}

\date{\today}

\begin{abstract}
The quantum efficiency of an electroluminescent intersubband
emitter based on InAs/AlSb has been measured as a function of the
magnetic field up to 20T. Two series of oscillations periodic in
1/B are observed, corresponding to the elastic and inelastic
scattering of electrons of the upper state of the radiative
transitions. Experimental results are accurately reproduced by a
calculation of the excited state lifetime as a function of the
applied magnetic field. The interpretation of these data gives an
exact measure of the relative weight of the scattering mechanisms
and allows the extraction of material parameters such as the
energy dependent electron effective mass and the optical phonon
energy.
\end{abstract}

\maketitle

The very short subband lifetime, generally in the order of 1ps, is
one of the major limitation to the optical gain of mid-infrared
(MIR) quantum cascade (QC) lasers\cite{Faist94}. These unipolar
devices are the only semiconductor lasers operating in continuous
wave up to room temperature\cite{Beck02,Faugeras05} in the 4 -
12$\mu$m wavelength range and have a great potential for future
applications involving molecular spectroscopy. The understanding
of the fundamental limits controlling the upper state lifetime
will be very beneficial for the ultimate design of these quantum
devices.

It is generally assumed that the short lifetime of electrons on
excited subbands is totally controlled by the interaction of
electrons with optical phonons. However, this scattering mechanism
is not the only factor which contributes to the lifetime. Indeed,
it has been recently demonstrated that other well known
fundamental elastic diffusive phenomena such as interface
roughness and alloy disorder can have scattering times of the same
order of that induced by the electron-phonon interaction
\cite{Leuliet05}, thus clearly contributing to the final lifetime
of electrons on excited subbands. These studies have been
conducted by applying a strong magnetic field parallel to the
growth axis, which breaks the in-plane free particle-like
dispersion and reduces the dimensionality of the carrier motion.
By tuning the magnetic field intensity it is possible to control
the excited state lifetime of an intersubband transition by
inhibiting or enhancing the elastic and inelastic scattering
resonances. This leads to oscillations of the excited state
lifetime and, as a consequence, of both the light intensity and
the electrical characteristics of the
device\cite{Smirnov02,Smirnov02a,Becker02}.

Previous works have been concentrating on MIR QC lasers. The
optical power oscillations as a function of the magnetic field,
P(B), reveal the presence of both the inelastic and the elastic
scattering of carriers\cite{Leuliet05}. Because these oscillations
are quasi-periodic in 1/B, an electron scattering spectroscopy
becomes possible by performing a Fourrier analysis of P(1/B),
which provides the characteristic energies involved in the
scattering mechanisms.  Until today the electron scattering
spectroscopy in high magnetic fields has been performed on QC
lasers only. In the context of a laser structure the information
that can be extracted are related to the variation of the
population inversion, therefore to the difference between the
lifetime of the upper state and that of the lower state of the
laser transition, $\tau_{2}-\tau_{1}$. To get direct information
solely on the lifetime of electrons on excited subbands one should
examine an electroluminescent structure. In this case the light
intensity is directly proportional to the population of electrons
on the upper state of the radiative transition, thus to its
lifetime.

In this article, we present a scattering spectroscopy
investigation of a two-level electroluminescent structure
specially designed to obtain information about the fundamental
mechanisms that control the lifetime of an excited subband in the
InAs/AlSb material system. This material system has recently been
exploited to produce lasers in the 3-5$\mu$m wavelength range of
energy with room temperature laser emission\cite{Teissier04},
inter-subband emission with a wavelength down to 2.5$\mu$m has
also been reported\cite{Barate05}. Moreover, this heterostructure
is composed by binary materials only and therefore the alloy
scattering cannot contribute to the upper state lifetime. Our
experiment shows also that it is possible to perform scattering
spectroscopy on weak electroluminescence signal from a MIR QC
structure.

\begin{figure}
\includegraphics[width=0.65\linewidth,angle=270,clip]{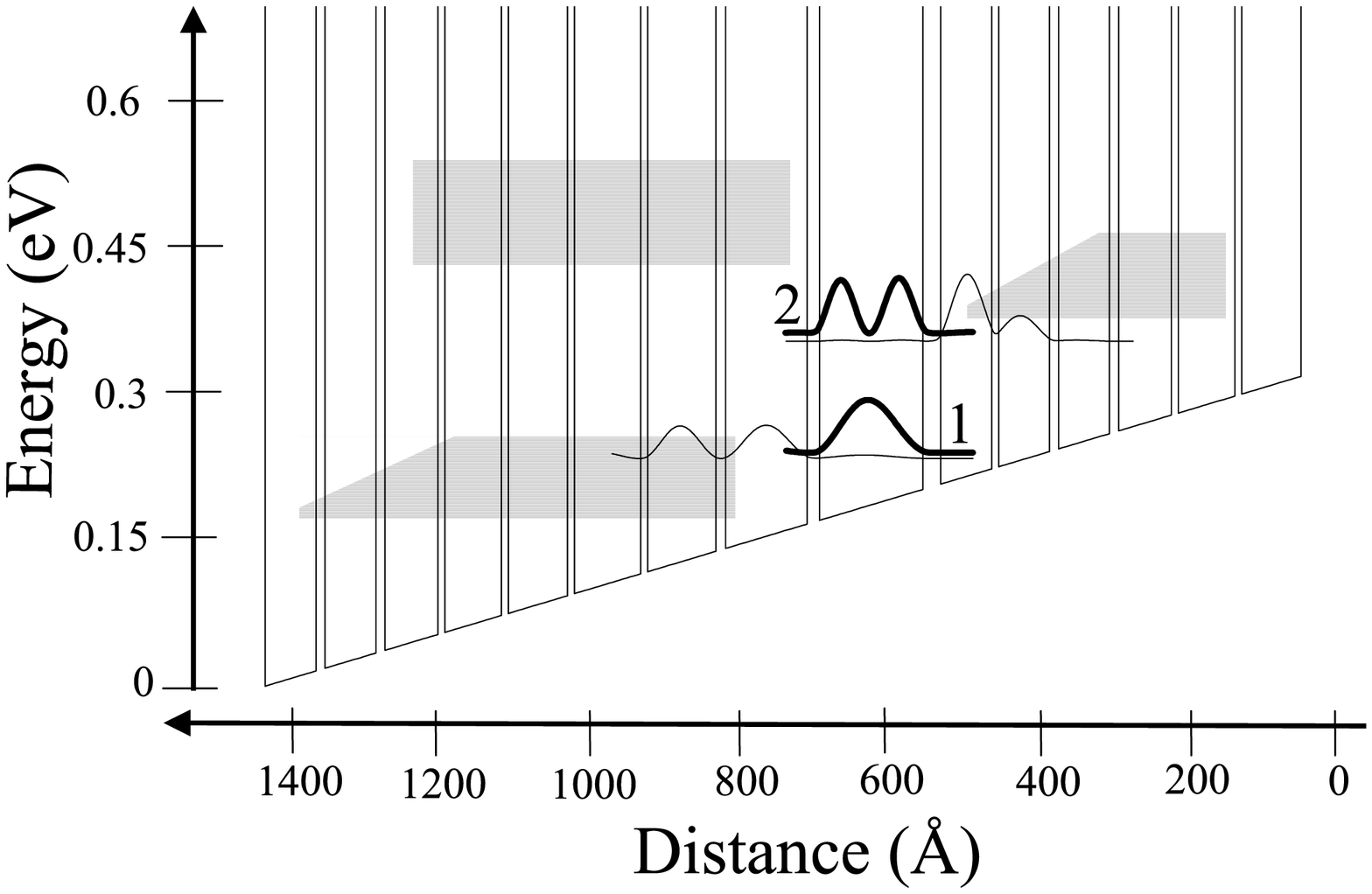}
\caption{\label{struc} Conduction band diagram of one period of
the active region of the two levels quantum cascade structure. The
radiative transition involves the first two levels in the central
quantum well (thick solid lines labelled 1 and 2). The gray blocks
indicate some of the superlattice minibands. From the injection
barrier, the layer sequence in angstroms is (bold layers are AlSb,
roman layer are InAs, doped layers are underlined):
\textbf{24}/132/\textbf{16}/104/\textbf{14}/90/\textbf{10}/84/
\textbf{10}/76/\textbf{10}/\underline{72}/\textbf{10}/\underline{68}/\textbf{12}
/\underline{65}/\textbf{12}/\underline{65}/\textbf{10}/65.}
\end{figure}

Fig.~\ref{struc} shows the conduction band diagram of the
two-level InAs/AlSb active region for an applied electric field of
22 kV/cm. It is composed of an injector miniband injecting
electrons into the second level of a quantum well (state labelled
2 in Fig.~\ref{struc}). Electrons can then be scattered on the
first subband of this quantum well (state labelled 1 in
Fig.~\ref{struc}) either in a radiative way, or by elastic or
inelastic scattering. The active region is composed of 10 periods
in series. For electrical and optical characterization,
non-alloyed Cr-Au contacts were evaporated and the samples were
processed into square mesas of $80\mu m \times 80 \mu m$. The
substrate has been polished at an angle of 45$^\circ$ with respect
to the growth axis to collect the luminescence. Measurements have
been performed in pulsed operation with pulses of 70 $\mu$s at a
repetition rate of 3.11 kHz.

\begin{figure}
\includegraphics[width=1.0\linewidth,angle=0,clip]{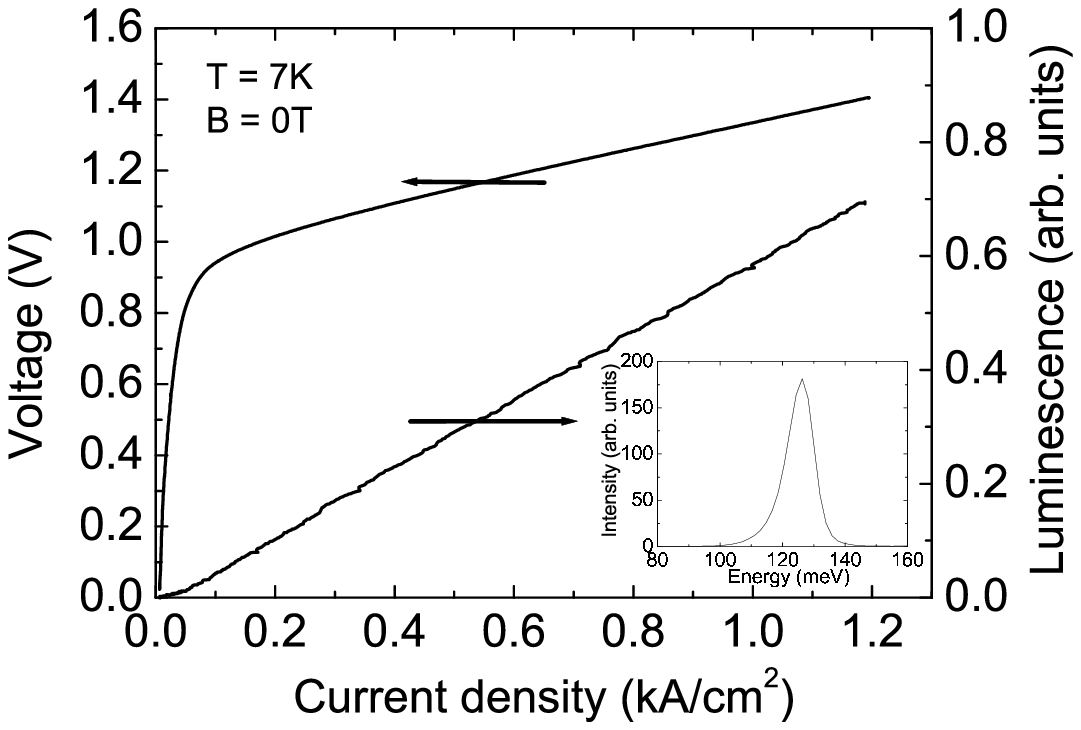}
\caption{\label{LIV} Voltage and electroluminescence intensity as
a function of the injected current density in pulsed operation (77
$\mu s$ at 3.11 kHz) in a 80 $\times$ 80 $\mu m^2$ square mesa at
T = 7K. Inset: Emission spectrum measured at 77 K.}
\end{figure}

The electrical and optical characteristics of this structure are
summarized in Fig.~\ref{LIV} which shows the evolution of the
voltage and of the electroluminescence measured with a blocked
impurity band (BIB) detector as a function of the injected current
density at a temperature of 7K and at $B = 0T$. For voltages
higher than 1 V, the different periods of the structure are
aligned and thus electrons can tunnel from the injector into the
excited state of the quantum well. The luminescence shows a well
defined linear dependance with the injected current, showing that
the quantum efficiency, at a fixed temperature, is constant with
the injected current. As can be seen in the inset of
Fig.~\ref{LIV}, the emission spectrum measured with a Nicolet
Fourier Transform Spectrometer is centered at 126 meV with a full
width at half maximum of 10 meV.

\begin{figure}
\includegraphics[width=1.0\linewidth,angle=0,clip]{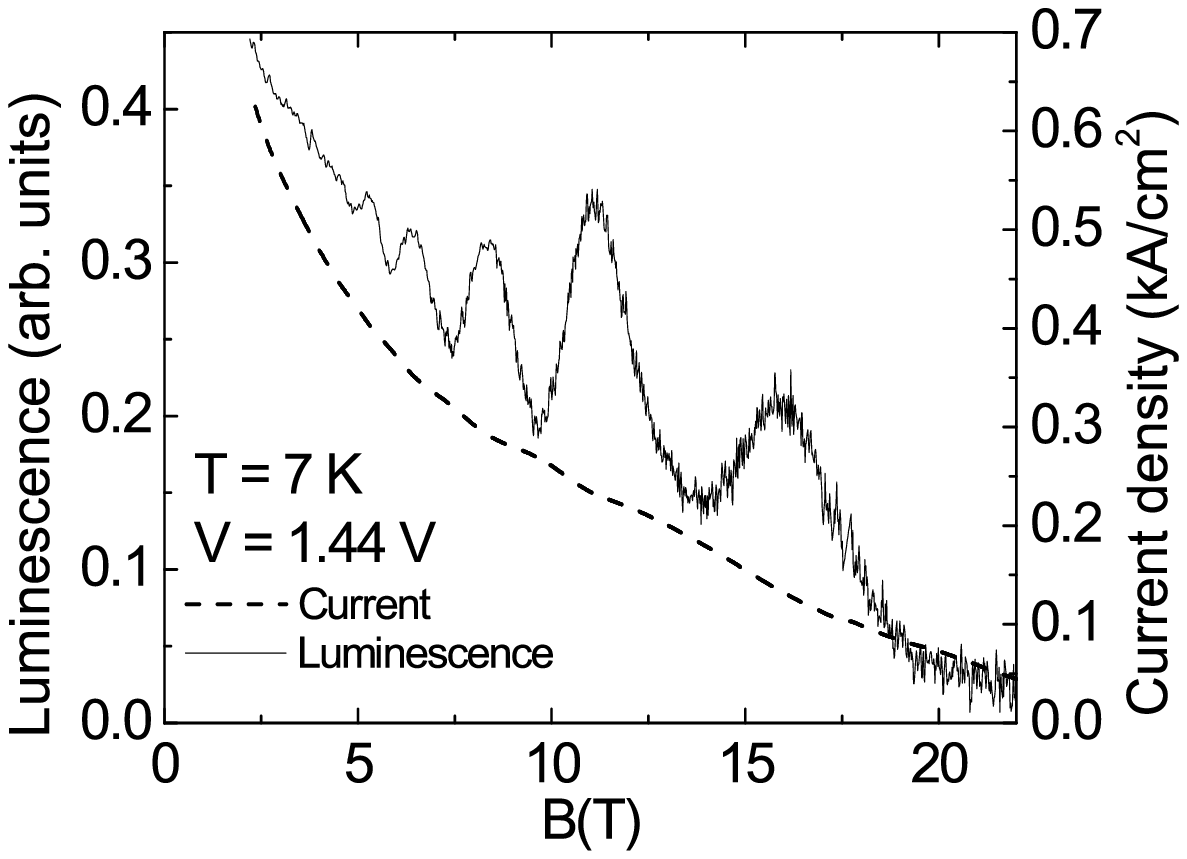}
\caption{\label{ILB} Evolution of the injected current (dashed
line) and of the luminescence (solid line) as a function of the
magnetic field at a fixed voltage of V = 1.44 V in pulsed
operation.}
\end{figure}

The experiment we have performed consists of fixing the voltage at
a constant value above the alignment voltage and then measuring
the evolution of the injected current and the luminescence
intensity as a function of the magnetic field applied parallel to
the growth axis. The results of this experiment for a constant
voltage of V = 1.44 V are presented in Fig.~\ref{ILB} for magnetic
fields up to 22 T. With increasing magnetic fields, two effects
are observed. The first effect is a strong decrease of the
injected current and as a consequence, of the light intensity. The
second effect is a series of oscillations in the injected current
and the luminescence intensity superimposed to the decreasing
background.

The strong decrease of the current as a function of the external
magnetic field is probably due to the magneto resistance that
appears in the two contacts and that reduces the voltage applied
to the active region of the order of $25 \%$ between $B=0T$ and
$B=20T$. For our QCS, the energy of the transition doesn't change
within $1\%$ of this variation of the voltage and only the
injection efficiency is affected. We have performed experiments on
similar two-level QCS but elaborated in other material systems
that showed the same behavior. For values of the magnetic field
above 20T, the current injected into the structure becomes smaller
than 5 mA and the luminescence signal can no longer be detected.

\begin{figure}
\includegraphics[width=1\linewidth,angle=0,clip]{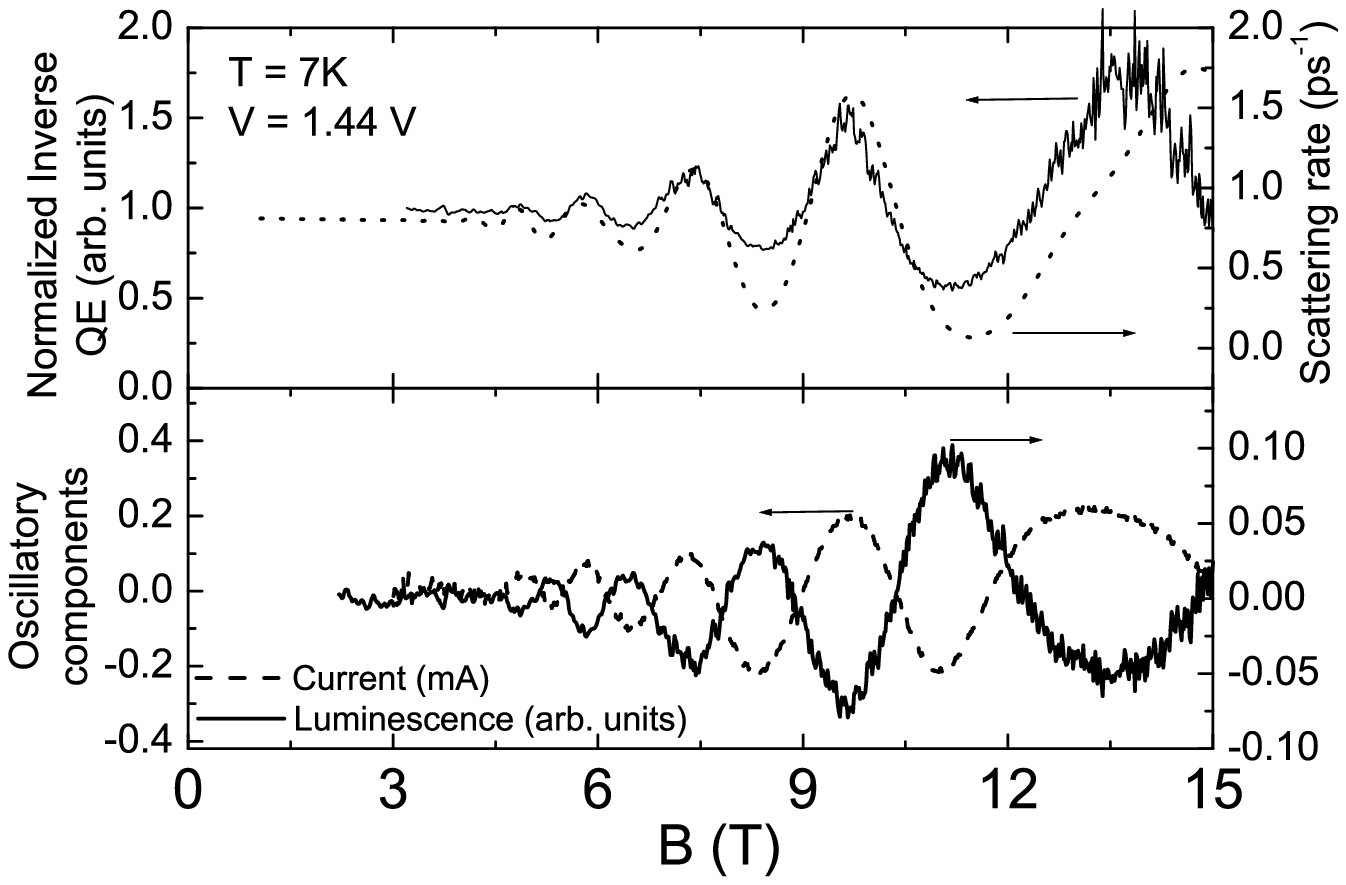}
\caption{\label{QE} Upper panel: Evolution of the inverse quantum
efficiency (solid line) and of the calculated excited state
scattering rate $\tau_2$ (dashed line) as a function of the
magnetic field at a fixed voltage of V = 1.44 V. Lower panel:
Oscillatory component of the injected current (dashed line) and of
the luminescence intensity (solid line) as a function of the
magnetic field.}
\end{figure}

The oscillations observed in both current and luminescence
intensity as a function of the magnetic field can be extracted
from the raw data by substracting a smoothed background. The
oscillatory components of both the current and the luminescence
are presented in the lower panel of Fig.~\ref{QE}. These
oscillations are interpreted as revealing the magnetic field
modulation of the excited state lifetime. They are in anti-phase
and can be explained as follows: When a Landau level of the
fundamental state $\vert1,N\rangle$ (where 1 refers to the
electronic subband and N is the Landau level index), is in
resonance with the first Landau level of the excited state
$\vert2,0\rangle$, electrons can be elastically scattered into the
state $\vert1,N\rangle$ which has no oscillator strength with
$\vert1,0\rangle$. Therefore, the electroluminescence decreases.
On the contrary, the current increases since the number of
available states for the tunnelling has strongly increased. In
fact, electrons can tunnel not only from the injector into
$\vert2,0\rangle$ but also into $\vert1,N\rangle$. When the
magnetic field is such that a Landau level $\vert1,N\rangle$ of
the fundamental state of the quantum well is one LO phonon energy
below the first Landau level of the excited state
$\vert2,0\rangle$, electrons in the excited state can emit
resonantly LO phonons and are efficiently scattered. This
intersubband-magneto-phonon effect, first observed in double
barrier resonant tunnelling diodes \cite{Eaves88} and then in QCLs
\cite{Smirnov02a}, leads to a decrease of the luminescence caused
by a decrease of the excited state lifetime in resonant condition.
Resonant magnetic fields verify the relation:
\begin{equation}
E_2-E_1-p\hbar\omega_{LO}=N\hbar\frac{eB}{m^*(E')} \label{IMPE}
\end{equation}
where $E_2-E_1$ is the intersubband energy separation which can be
measured, $\hbar\omega_{LO}$ is the LO phonon energy, $B$ is the
magnetic field, $m^*(E)$ is the energy dependent electron
effective mass, N is an integer, $p =0$ or $1$ for the elastic and
inelastic series respectively and $E'=E_{2}-p\hbar\omega_{LO}$. We
recall that these oscillations directly reveal the different
electron scattering mechanisms, whereas in QC lasers, the observed
oscillations reveal variations in the population inversion ruling
laser action.

\begin{figure}
\includegraphics[width=0.6\linewidth,angle=270,clip]{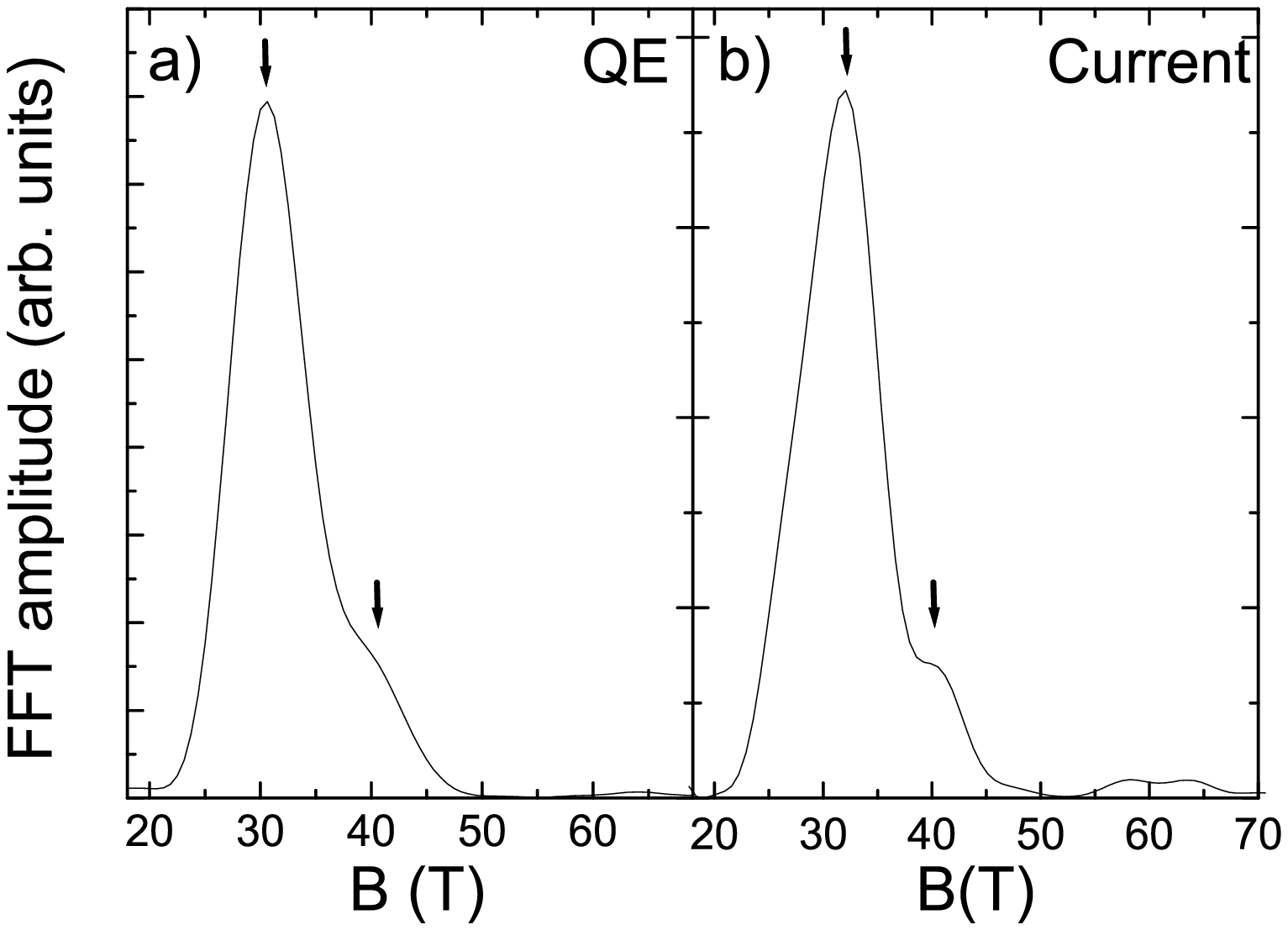}
\caption{\label{FFT} Fourier transform amplitude in 1/B of a) the
quantum efficiency and b) the current. In both cases, two
frequencies are extracted $B_{F1} = 30 T$ and $B_{F2} = 40 T$.}
\end{figure}

In order to avoid any artifacts coming from the variations of the
current injection with magnetic field, we have determined the
quantum efficiency (QE) of our electroluminescent structure by
dividing the luminescence signal by the injected current. The
evolution of the inverse quantum efficiency as a function of the
magnetic field is presented in the upper panel of Fig.~\ref{QE}
(solid line). At $B = 0T$, this quantity is constant as a function
of the current as is shown by the linear behavior of the
luminescence in Fig.~\ref{LIV}. When a magnetic field is applied,
the quantum efficiency oscillates with B. A Fourier analysis of
these oscillations expressed in inverse magnetic field is
presented in Fig.~\ref{FFT} a). Two distinct series are evident, a
well defined series with a fundamental field of $B_{F1}=30 T$ and
a second series with a lower intensity and with a fundamental
field $B_{F2}=40 T$. A similar Fourier analysis of the
oscillations of the current is presented in Fig.~\ref{FFT} b) and
shows exactly the same characteristic frequencies. These two peaks
reveal both the LO phonon inelastic scattering mechanism for the
low field series and the elastic scattering for the high field
series.

In a simple rate equation model considering a reservoir state
injecting electrons with an efficiency $\eta$ in the excited state
2, the quantum efficiency of a period is defined by:
\begin{equation}
QE=\frac{P}{J}\approx\frac{\eta}{q}\frac{\tau_2}{\tau_{rad}}
\label{QEmat}
\end{equation}

where $P$ is the optical power, $J$ the current density, $q$ the
electrical charge, $\eta$ is the injection efficiency,
$\tau_{rad}$ is the radiative lifetime from level 2 to level 1
(constant with $B$), $\tau_2$ is the excited state lifetime given
by $\tau_2^{-1}=(\tau_2^{elastic})^{-1}+(\tau_2^{LO})^{-1}$ where
$\tau_2^{elastic}$ is the elastic scattering time and
$\tau_2^{LO}$ is the LO phonon scattering time.

Because it is the quantity that we calculate directly, we present
in the upper panel of Fig.~\ref{QE} the evolution of the
calculated excited state scattering rate for our structure (black
dashed line) compared to the measured inverse quantum efficiency
(black solid line). Experimental results are well reproduced by a
calculation considering electron-LO phonon scattering and
interface roughness scattering. This calculation is presented in
details in Ref.\cite{Leuliet05} for quantum cascade lasers.

We have been able to reproduce the observed oscillations with a
phonon energy of 30.4 meV, in good agreement with values reported
in the literature \cite{Groenen98}. We have used a Landau level
broadening of 6 meV and an energy dependent effective mass taking
non-parabolicity into account through a 3 band $\textbf{k} \cdot
\textbf{p}$ model. The parameters used in the 3 band $\textbf{k}
\cdot \textbf{p}$ calculation are a band gap of 0.417 eV, a
split-off band energy of 0.390 eV and a conduction band effective
mass at $\Gamma$ of $m^{*}=0.023 m_{0}$ where $m_0$ is the
electron mass in vacuum. At $B =0 T$, the effective mass of the
excited state deduced from our calculation is $0.035 m_0$. At low
temperatures, the main contribution to the elastic series is the
interface roughness scattering \cite{Leuliet05}. For our
calculation, we have used a gaussian autocorrelation of the
interface roughness with a correlation length of $60 \AA$ and an
average height of the roughness of $2.0 \AA$, slightly lower than
values reported so far for this material system
\cite{Bolognesi92}. The interface roughness in this material
system is of the same amount as the one found in GaAs/AlGaAs
interfaces \cite{Leuliet05}. The deduced scattering times at
$B=0T$ are $\tau_{LO} = 1.57 ps$ for the optical phonon scattering
and $\tau_{IR} =6.78 ps$ for the interface roughness scattering.
For magnetic fields above 15 T, we observe discrepancies between
experimental results and the calculation which we think are a
consequence of the decrease of the voltage applied to the
structure with increasing magnetic fields.

In conclusion, we have reported on the first direct measure of the
magnetic field dependence of the quantum efficiency of a two-level
QCS up to 20 T. This quantum efficiency oscillates as a function
of the magnetic field. Two distinct series of oscillations are
identified: the inelastic LO phonon scattering mechanism and, of
much smaller intensity, the elastic scattering due to the
interface roughness. Experimental data are well reproduced by a
calculation of the excited state lifetime considering electron-LO
phonon and interface roughness scattering processes. Two-level QCS
offers an ideal system to study the electron scattering mechanisms
in QCS and to determine their relative importance in QCS of
different material systems.

\begin{acknowledgments}
We gratefully acknowledge support from EU FP6 Grant STRP 505642
"ANSWER", from EU MRTN-CT-2004-51240 "POISE" and from NHMFL In
House Research Program (project 5053).
\end{acknowledgments}

\bibliography{}

\end{document}